\documentclass[10pt]{article}
\usepackage{graphicx}
\usepackage{lineno}

\def\Title#1{\begin{center} {\Large #1 } \end{center}}
\def\Author#1{\begin{center}{ \sc #1} \end{center}}
\def\Address#1{\begin{center}{ \it #1} \end{center}}

\newcommand\pubblock{\rightline{\begin{tabular}{l} Proceedings of the Fifth Annual LHCP\\ \pubnumber\\
         \pubdate  \end{tabular}}}

\newenvironment{Abstract}{\begin{quotation} \begin{center} 
             \large ABSTRACT \end{center}\bigskip 
      \begin{center}\begin{large}}{\end{large}\end{center} \end{quotation}}

\newenvironment{Presented}{\begin{quotation} \begin{center} 
             PRESENTED AT\end{center}\bigskip 
      \begin{center}\begin{large}}{\end{large}\end{center} \end{quotation}}

\def\beq{\begin{equation}}
\def\eeq#1{\label{#1}\end{equation}}
\def\eeqn{\end{equation}}

\def\beqa{\begin{eqnarray}}
\def\eeqa#1{\label{#1}\end{eqnarray}}
\def\eeqan{\end{eqnarray}}

\def\Wp{\ensuremath{W^{+}}}
\def\Wm{\ensuremath{W^{-}}}
\def\W{\ensuremath{W}}
\def\Z{\ensuremath{Z}}
\def\gammastar{\ensuremath{\gamma^{*}}}

\def\Wenu{\ensuremath{W\rightarrow e\nu}}
\def\Wlnu{\ensuremath{W\rightarrow\ell\nu}}
\def\Zll{\ensuremath{Z/\gammastar\rightarrow\ell\ell}}

\def\Zee{\ensuremath{Z/\gammastar\rightarrow ee}}
\def\pTl{\ensuremath{p_{\mathrm{T}}^{\ell}}}
\def\etalep{\ensuremath{\eta^{\ell}}}
\def\etae{\ensuremath{\eta^{e}}}
\def\pTnu{\ensuremath{p_{\mathrm{T}}^{\nu}}}
\def\mT{\ensuremath{m_{\mathrm{T}}}}
\def\mll{\ensuremath{m_{\ell\ell}}}
\def\yll{\ensuremath{y_{\ell\ell}}}
\def\Wplnu{\ensuremath{W^+\rightarrow\ell^+\nu}}
\def\chisqndf{\ensuremath{\chi^2/{\mathrm{ndf}}}}
\def\tt{\ensuremath{t\bar{t}}}
\def\alfas{\ensuremath{\alpha_S}}
\def\sqrts{\ensuremath{\sqrt{s}}}

\let\bar=\overbar

\def\Dslash{\not{\hbox{\kern-4pt $D$}}}
\def\dslash{\not{\hbox{\kern-2pt $\del$}}}

\def\msb{{\bar{\ssstyle M \kern -1pt S}}}

\usepackage{color}

\newcommand\pubnumber{ ATL-PHYS-PROC-2017-111 }

\newcommand\pubdate{\today}

\def\affiliation{
On behalf of the ATLAS Collaboration, \\
Santa Cruz Institute for Particle Physics, University of California Santa Cruz \\
Santa Cruz, CA 95064, US}

\begin{document}
\large
\begin{titlepage}
\pubblock

\vfill
\Title{  Single boson production and differential cross section measurements in ATLAS }
\vfill

\Author{ Chiara Debenedetti  }
\Address{\affiliation}
\vfill
\begin{Abstract}

High-precision measurements of the Drell-Yan production provide information on different aspects contributing to the process, such as parton distribution functions (PDFs), and comparisons with the current precision reached theoretically on the calculations of the cross sections of such processes.
This document describes ATLAS measurements, performed at different centre-of-mass energies, of vector boson (\W\ and \Z) cross sections and cross-section ratios.
It also discusses measurements of ratios of \Z-boson and top-quark pair production cross sections, which provide important information on the proton PDFs.

\end{Abstract}
\vfill

\begin{Presented}
The Fifth Annual Conference\\
 on Large Hadron Collider Physics \\
Shanghai Jiao Tong University, Shanghai, China\\ 
May 15-20, 2017
\end{Presented}
\vfill
\end{titlepage}
\def\thefootnote{\fnsymbol{footnote}}
\setcounter{footnote}{0}
%

\normalsize 


\section{Introduction}

Calculations of the Drell-Yan production process are known at next-to-next-to-leading order (NNLO) precision in QCD and next-to-leading order (NLO) precision in electroweak.
The high production rate of such processes at the LHC leads to very high experimental precision in their measurement, necessary to probe and challenge the available predictions.
These measurements provide information on all the different contributions to the process, such as parton distribution functions (PDFs).
In the latest FEWZ and DYNNLO calculations~\cite{Gavin:2010az,Gavin:2012sy,Li:2012wna,Catani:2007vq,Catani:2009sm}, the largest uncertainties come from renormalisation and factorisation scales ($\sim1.1\%$), and from PDFs ($\sim 2.5\%$).
This document describes ATLAS~\cite{Aad:2008zzm} measurements, performed at different centre-of-mass energies, of vector boson (\W\ and \Z) cross sections and cross-section ratios, and ratios of \Z-boson and top-quark pair production cross sections.

\section{Precision measurement and interpretation of inclusive \Wp, \Wm\ and \Z/\gammastar production cross sections with the ATLAS detector}

The most precise measurement of \W and \Z-boson cross sections in the leptonic decay channels was performed using data collected at $\sqrt{s}=7$~GeV with the ATLAS detector, corresponding to an integrated luminosity of $4.7$~fb$^{-1}$~\cite{Aaboud:2016btc}.
The analysis cuts select $\sim 30$~M \W\ and $\sim 3$~M \Z-boson events.
The analysis is performed separately on electron and muon final states,
 which are later combined.
The fiducial selection is similar to the one performed on reconstructed events.
The \W\ analysis fiducial selection cuts are: $\pTl>25$~GeV, $|\etalep |<2.5$, $\pTnu>25$~GeV, and $\mT>40$~GeV, where \mT\ is defined as follows: $\mT=\sqrt{2\pTl\pTnu[1-\cos{(\phi_{\ell}-\phi_{\nu})}]}$.
The \Z\ analysis selection, instead, defines different categories based on the lepton pseudorapidity, referred to as central-central (CC), and central-forward (CF).
The former exists for both the muon and electron decay channels, and refers to both leptons being within the region of the tracker acceptance, while the latter exists only for the electron case, and considers one lepton within the tracker acceptance, and one reconstructed using the calorimeter system in the forward region.
The fiducial selection cuts for the \Z\ channel are: $\pTl>20$~GeV, $|\etalep_1|<2.5$, $|\etalep_2|<2.5$ in the CC category, $2.5<|\etae_2|<4.9$ in the CF category.
%
%
%


The backgrounds
are rather small in both channels, amounting to about 8\% of the selected events in the \Wlnu\ case, to less than 1\% in the \Zll\ CC selection, and to about 3\% in the \Zee\ CF case.
All the backgrounds are evaluated using Monte Carlo simulation, except the multijet background, originating from QCD events faking leptons and missing energy.
This background, given the difficulty to describe fakes in the Monte Carlo simulation, is extracted using data-driven techniques.

The measurement is performed differentially for both channels.
In the \Wlnu\ analysis, ten bins of $|\etalep|$ are considered, to provide useful information on the initial state PDFs. 
To extract analogous information from the \Zll\ analysis, three slices in di-lepton invariant mass (\mll) are considered: $46<\mll/$GeV$<66$, $66<\mll/$GeV$<116$, and $116<\mll/$GeV$<150$, categorised in bins of \Z-boson rapidity (\yll): $0<|\yll|<2.4$ for the CC selection, and $1.6<|\yll|<3.6$ for the CF.


The main uncertainties in this analysis originate from the reconstruction efficiencies of the leptons, affecting the fiducial cross section calculation
 with a 1.51\% error in the \Zee\ CF analysis; from the matrix-element modelling of the \W-boson signal,
 introducing a 0.6\% error in the \Wenu\ analysis; from the uncertainty on the data driven method for the QCD-background estimate, similarly affecting the \Wenu\ channel, and from the luminosity measurement, affecting all channels with a 1.8\% error.
These uncertainties are nevertheless very small, and they are dramatically reduced with respect to the previous ATLAS measurement~\cite{Aad:2011dm}: the improvement on the lepton and missing energy uncertainty amounts to 60\%, and to 50\% on the luminosity uncertainty.

\begin{figure}[h]
\centering
\includegraphics[width=0.31\textwidth]{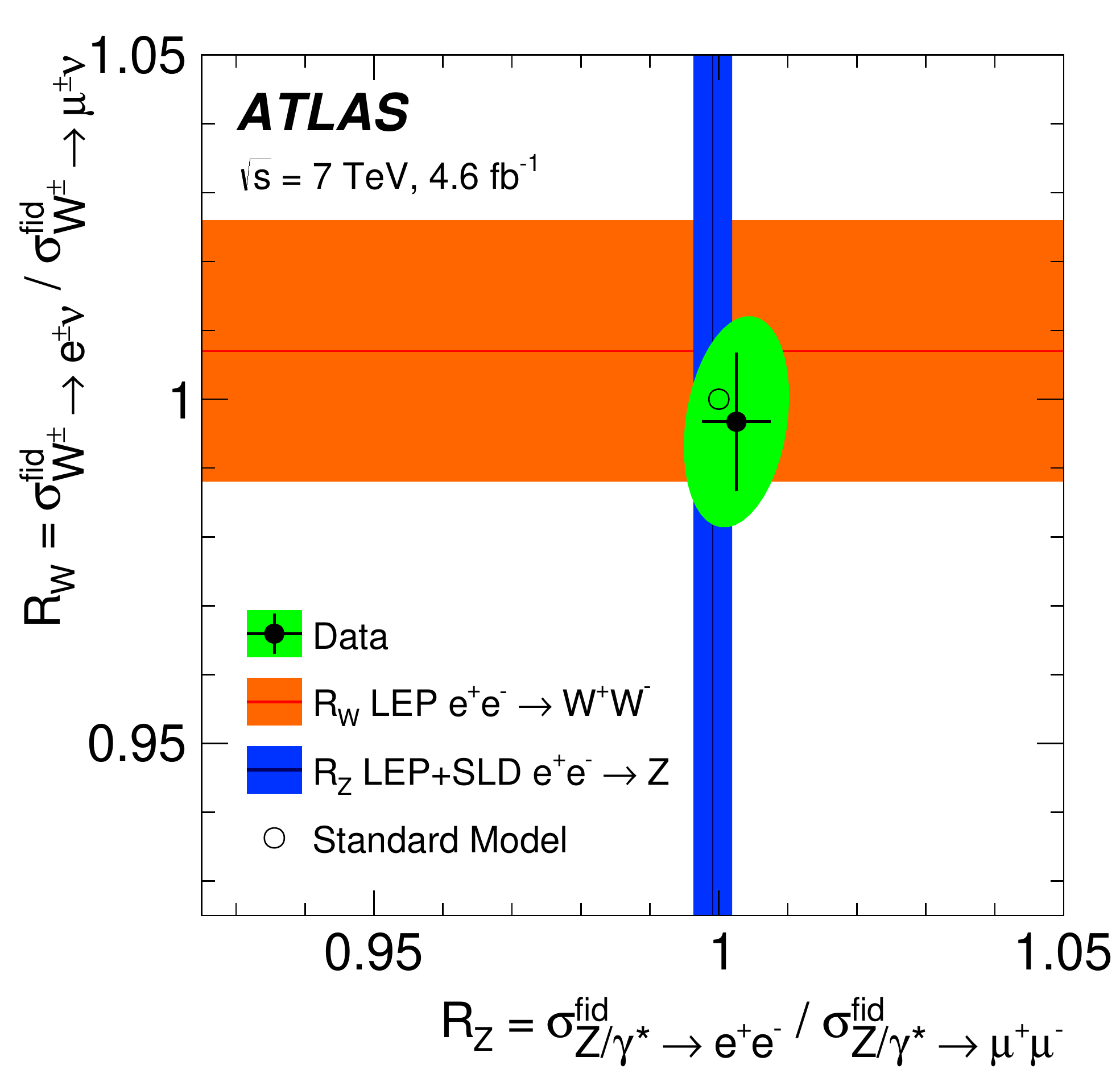}
\includegraphics[width=0.31\textwidth]{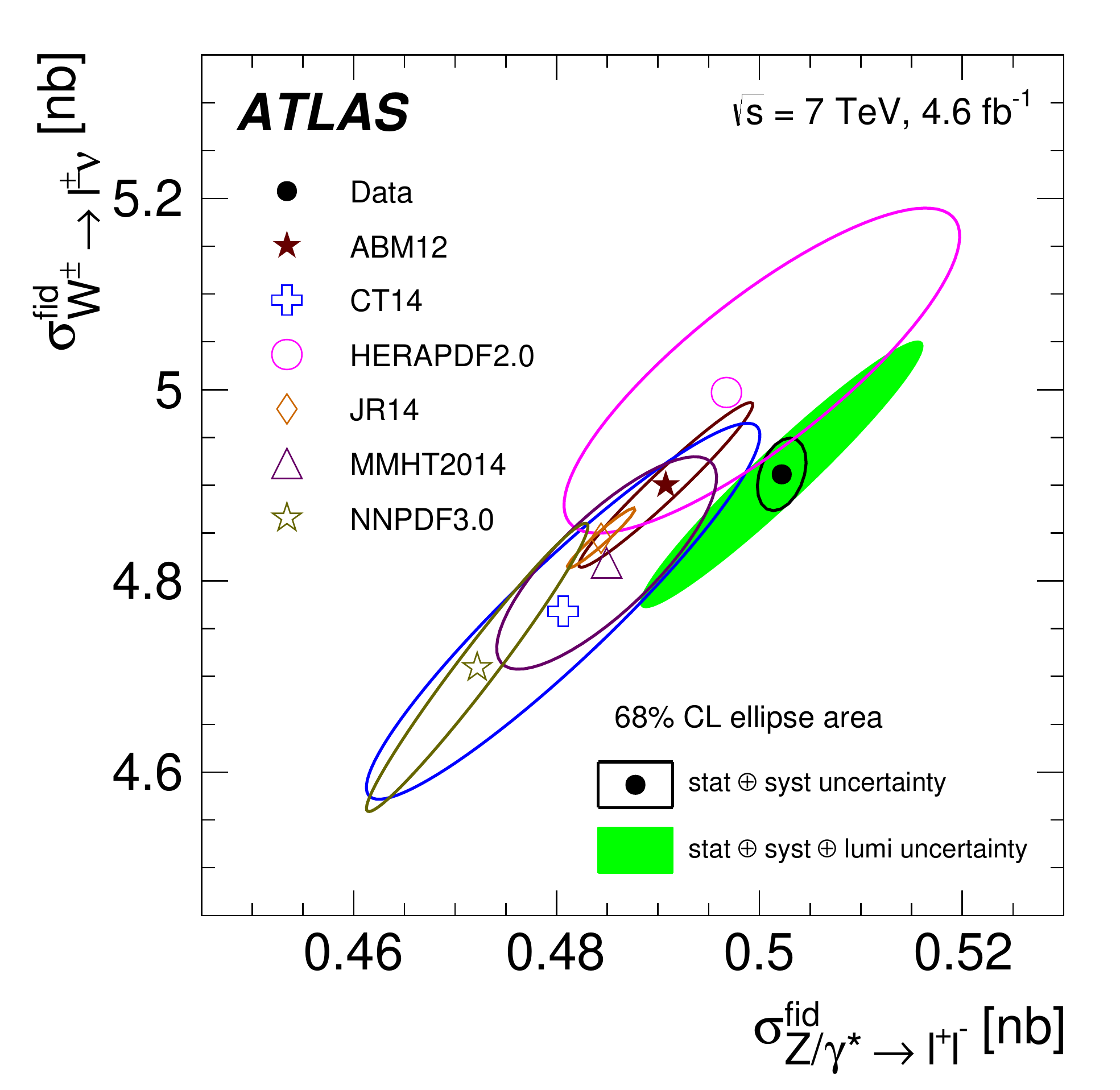}
\caption{The left plot represents a measurement of lepton universality, compared to the SM expectation. The orange and blue bands represent results from $e^+e^-$ colliders LEP and SLC~\cite{ALEPH:2005ab}. The green ellipse illustrates the 68\% CL for the correlated measurement of $R_{\W}$ and $R_{\Z}$, and the black error bars the standard deviation. The right plot shows the measured cross sections for \W\ vs \Z-boson. The data ellipses illustrate the 68\% CL coverage. Data are compared with predictions using different PDF sets, and their PDF uncertainties. Both plots are taken from Ref.~\cite{Aaboud:2016btc}.}
\label{fig:lep_universality_WZ}
\end{figure}

The fiducial cross sections are presented in Table~\ref{tab:fid_xsecs_WZ}, confirming lepton universality, shown in the left plot in Figure~\ref{fig:lep_universality_WZ}.
This measurement reaches a very high precision, of 1\% for $R_W$ and 0.5\% for $R_Z$, improving LEP results on on-shell \W-boson measurements~\cite{Schael:2013ita}, thanks to the cancellation of correlated uncertainties.
As can be seen in Table~\ref{tab:fid_xsecs_WZ}, the combination of the electron and muon channel achieves the best precision.

\begin{table}[h]
\begin{center}
\begin{tabular}{l|c|c}  
Lepton channel &$\sigma^{\mathrm{fid}}_{\Wlnu}$~[pb] &$\sigma^{\mathrm{fid}}_{\Zll}$~[pb] \\
\hline
electron &$4896\pm2\,(\mathrm{stat})\pm49\,(\mathrm{syst})\pm88\,(\mathrm{lumi})$ &$502.7\pm0.5\,(\mathrm{stat})\pm2.0\,(\mathrm{syst})\pm9.0\,(\mathrm{lumi})$ \\
muon &$4912\pm1\,(\mathrm{stat})\pm32\,(\mathrm{syst})\pm88\,(\mathrm{lumi})$ &$501.4\pm0.4\,(\mathrm{stat})\pm2.3\,(\mathrm{syst})\pm9.0\,(\mathrm{lumi})$ \\
combination &$4911\pm1\,(\mathrm{stat})\pm26\,(\mathrm{syst})\pm88\,(\mathrm{lumi})$ &$502.2\pm0.3\,(\mathrm{stat})\pm1.7\,(\mathrm{syst})\pm9.0\,(\mathrm{lumi})$ \\
\hline
\end{tabular}
\caption{ Measured fiducial cross sections with statistical and systematic uncertainties~\cite{Aaboud:2016btc}, for the \Wlnu\ and \Zll\ processes. The \Z\ cross section is calculated for $66<\mll/$GeV$<116$. The uncertainties denote the statistical (stat), the experimental systematic (syst), and the luminosity (lumi) components.}
\label{tab:fid_xsecs_WZ}
\end{center}
\end{table}

Given the very small uncertainty on the measurement, the result discriminates very effectively between different PDF sets, and a test of their compatibility with the measurement is shown in the right plot in Figure~\ref{fig:lep_universality_WZ}.
The compatibility with different PDF predictions is studied also via the calculation of the cross section ratios between the different charges of the \W-boson, and between \W- and \Z-boson. While the former is very well described by all the PDF sets, the latter is consistently larger than the calculations, hinting for an enhancement of the strange component in the PDFs.
This result is also compatible with another ATLAS measurement, performed on the data taken at the start of the 13 TeV run~\cite{Aad:2016naf}, although the latter includes larger uncertainties.

\begin{figure}[h]
\centering
\includegraphics[width=0.31\textwidth]{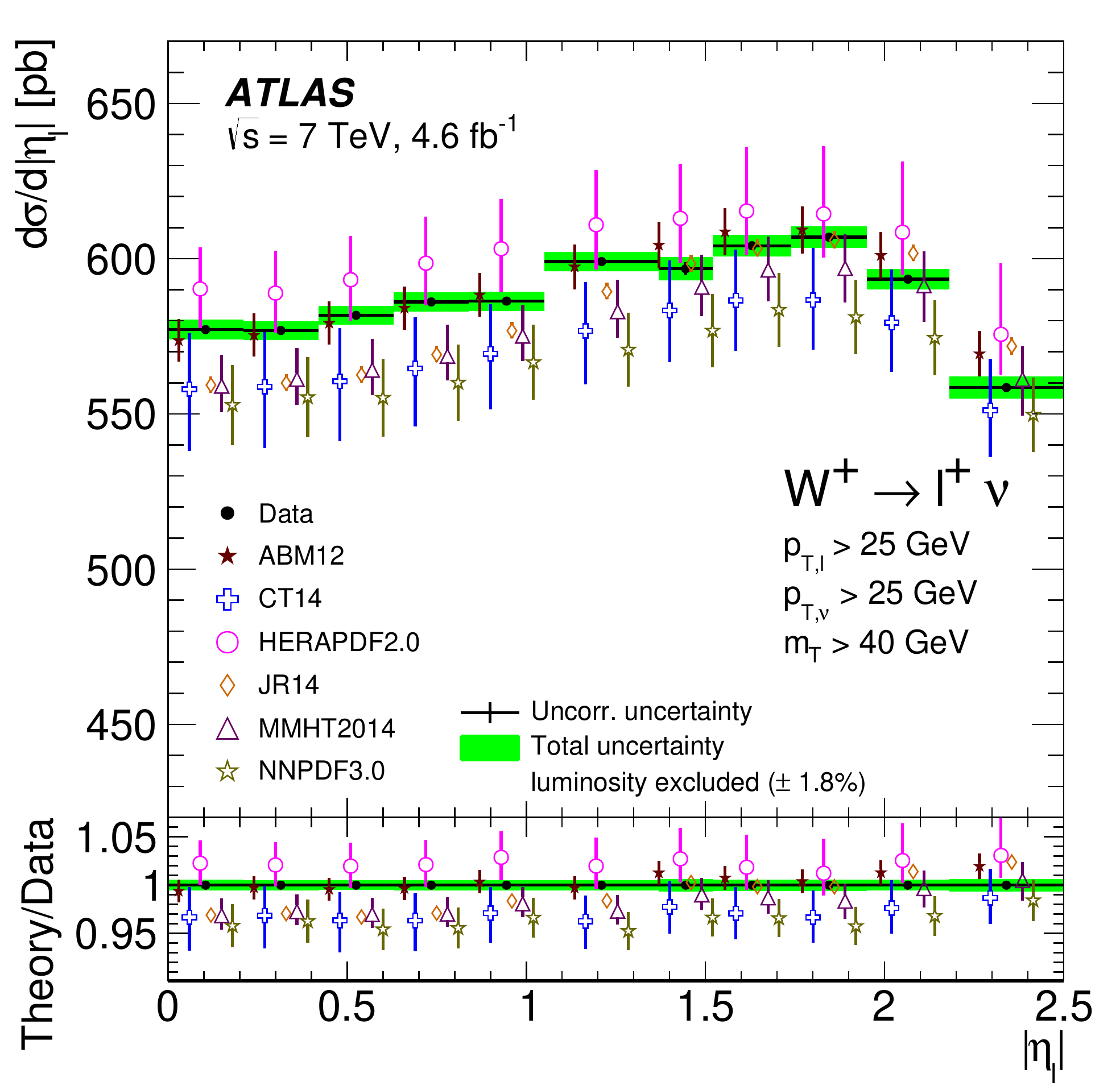}
\includegraphics[width=0.31\textwidth]{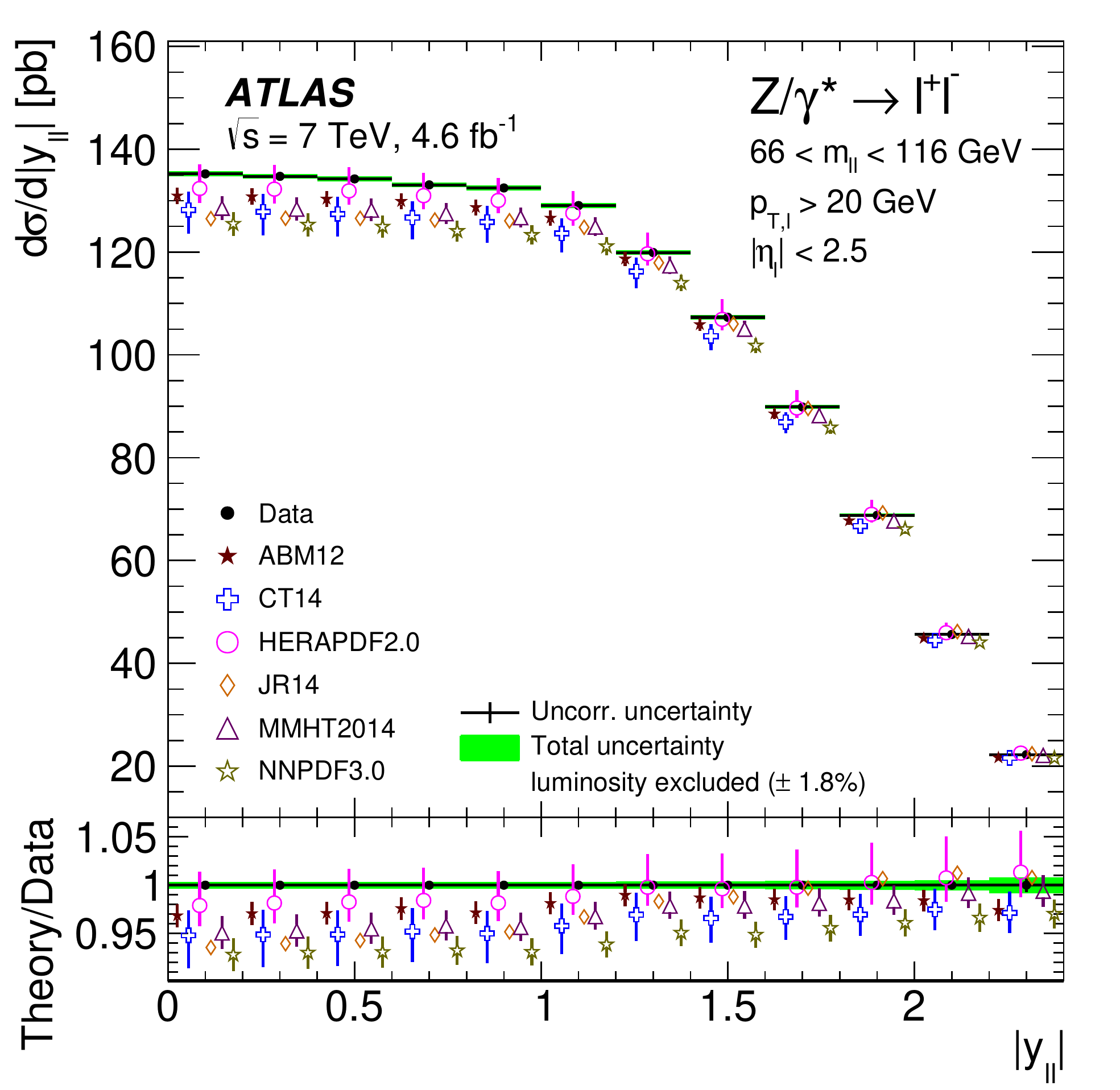}
\caption{The left plot shows the differential $\sigma$ measurement for \Wplnu. Predictions computed at NNLO QCD with NLO EW corrections using various PDF sets are compared to the data. The theory uncertainty corresponds to the quadratic sum of the PDF and the statistical uncertainties. The right plot shows instead the differential cross-section for \Zll\ in the \Z\ mass peak region, for central rapidity values. Both plots are taken from Ref.~\cite{Aaboud:2016btc}.}
\label{fig:diffXsecs_WZ}
\end{figure}

The combination between the two lepton channels is performed not only for the fiducial cross-section evaluation, but also for the differential case.
A $\chi^2$ fit is used to combine the two channels, and the compatibility between the two is found to be very good. 
The \W-boson differential cross section shapes are well described by the predictions, and only present normalisation differences, as shown in the left plot in Figure~\ref{fig:diffXsecs_WZ}, and also the description of the \W-boson charge asymmetry is excellently described by all the predictions.
A dependence in the \yll\ variable is instead observed when comparing data and predictions in the \Zll\ case, in the central rapidity region.
The predictions, in fact, underestimate the data up to 5\%, as can be observed in the right plot in Figure~\ref{fig:diffXsecs_WZ}.
The forward \yll\ region does not provide good discrimination power to be sensitive to the different PDF predictions.

\begin{figure}[h]
\centering
\includegraphics[width=0.31\textwidth]{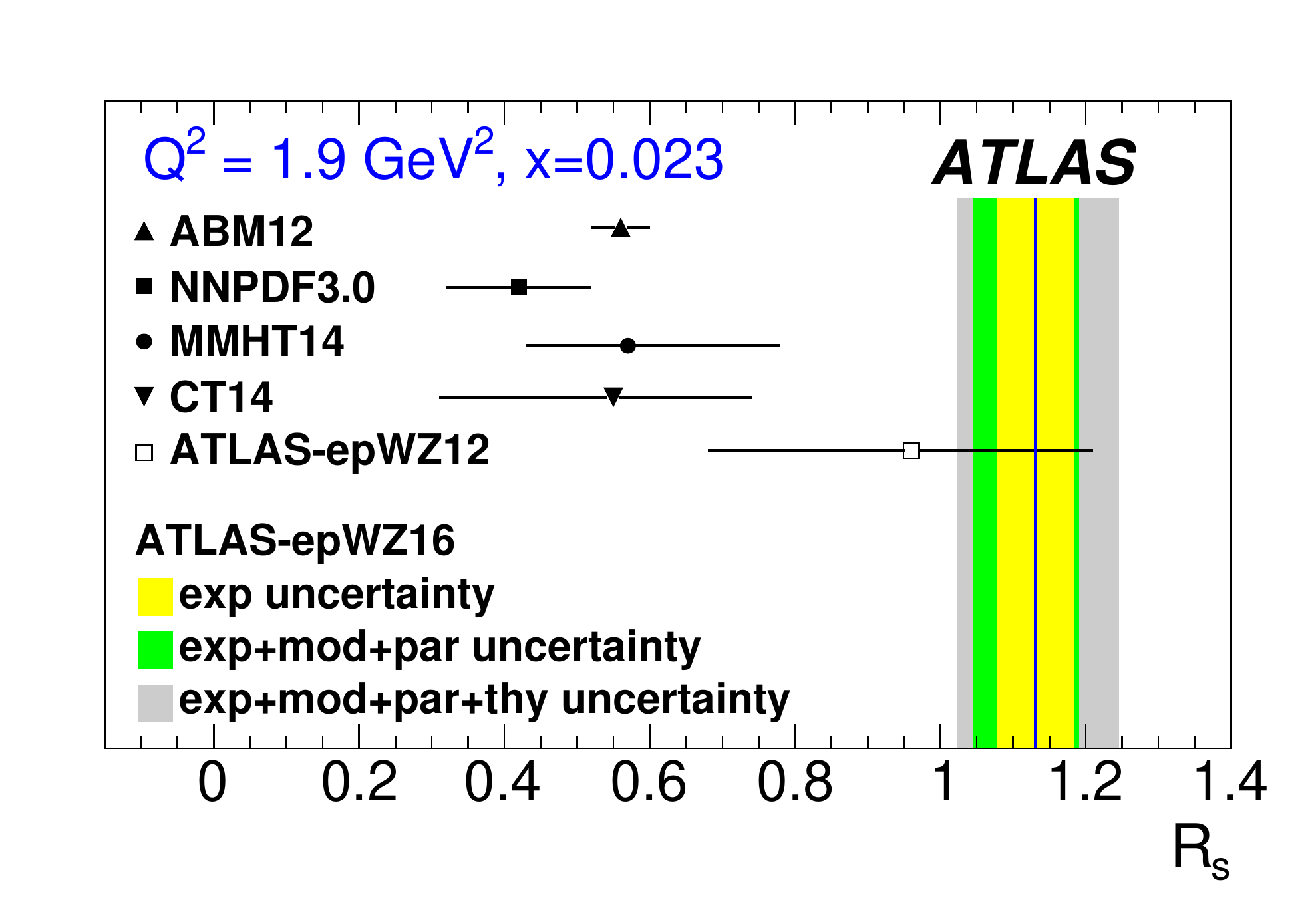}
\includegraphics[width=0.31\textwidth]{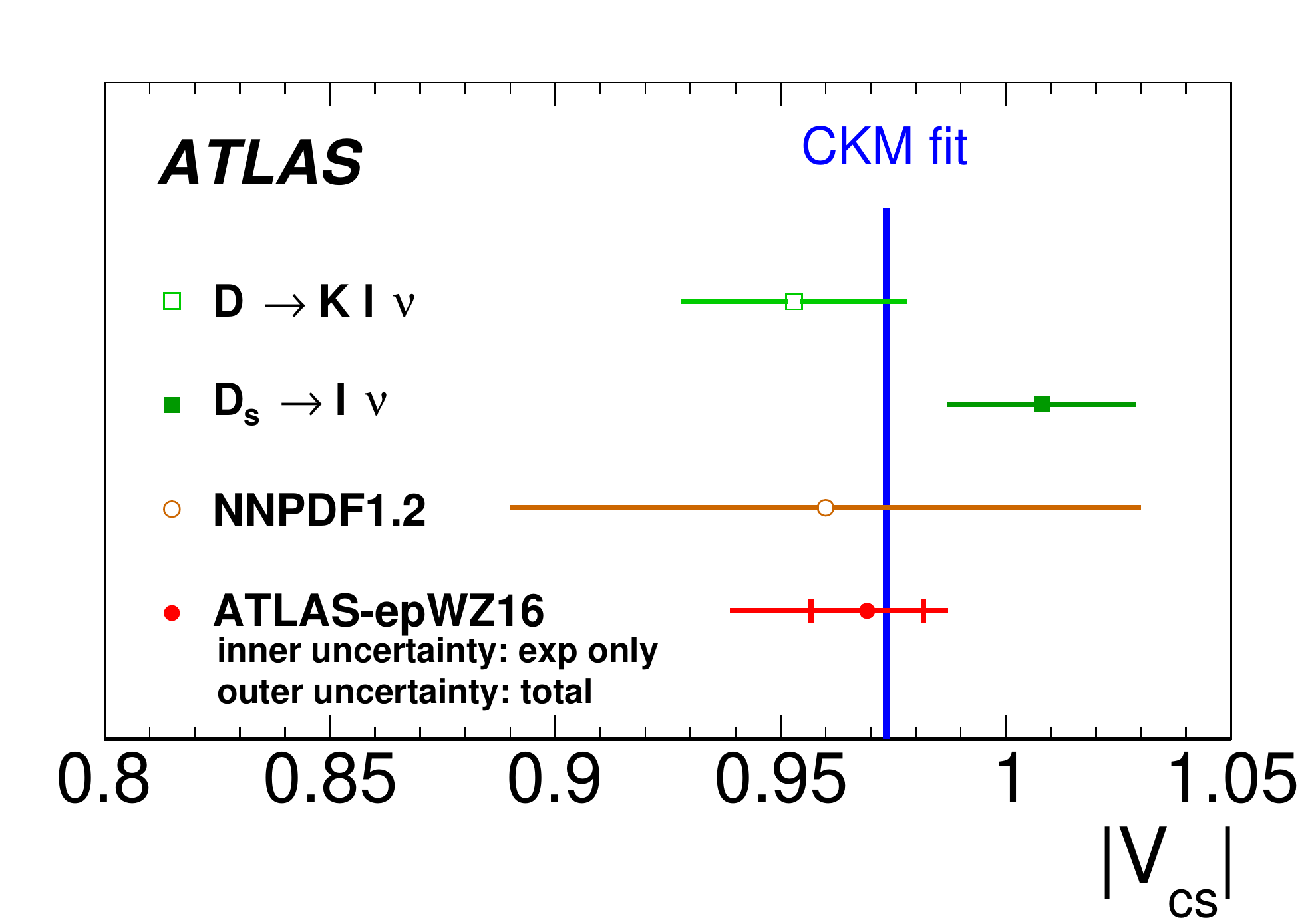}
\caption{The left plot shows the relative strange-to-down sea quark fraction $R_s$. The data line with its uncertainty contributions from experimental data, QCD fit, and theoretical uncertainties are compared to the predictions from different NNLO PDF sets. The right plot shows the $|V_{cs}|$ as determined in the global CKM fit cited by the PDG~\cite{Agashe:2014kda} (blue vertical line) compared to extractions from $D_s\rightarrow\ell\nu$ and $D\rightarrow K\ell\nu$~\cite{Agashe:2014kda} and the NNPDF1.2 fit~\cite{Ball:2009mk}. The ATLAS-epWZ16 fit result is shown with uncertainty contributions from the experimental data and the total uncertainty including all fit and further theoretical uncertainties. Both plots are taken from Ref.~\cite{Aaboud:2016btc}.}
\label{fig:Vcs_WZ}
\end{figure}

The measured data are further used to profile different PDF sets, and the one presenting the best agreement to the data is found to be the ATLAS-epWZ12 set~\cite{Aad:2012sb}, with \chisqndf=113/159.
The data provide constraints on both central values and uncertainties, significantly affecting the strange fraction, shifted to higher central values, confirming the enhancement hinted by the \W/\Z\ cross-section ratio measurement.
The effect of the strangeness enhancement can be seen in Figure~\ref{fig:Vcs_WZ}.
A new PDF set, ATLAS-epWZ16, is also obtained using the measured data, in combination with $ep$ data from the H1 and ZEUS experiments~\cite{Abramowicz:2015mha}.
The new data add information on the flavour composition of the quark sea and the low-$x$ valence quark distributions.
Furthermore, a measurement of the CKM $|V_{cs}|$ matrix element is performed, and found to be competitive with previous results, as shown in Figure~\ref{fig:Vcs_WZ}.


\section{Measurements of top-quark pair to \Z-boson cross-section ratios at $\sqrt{s}=13,8,7$~TeV with the ATLAS detector}

The described analysis is aimed at evaluating ratios and double ratios of \Z-boson and top-quark pair production cross sections obtained at different centre-of-mass energies, using the data collected by the ATLAS detector~\cite{Aaboud:2016zpd}.
The analysed quantities are:
\begin{itemize}
\item{$R^{\mathrm{fid}}_{Z_i/Z_j}=\sigma^{\mathrm{fid}}_{Z(i \mathrm{TeV})}/\sigma^{\mathrm{fid}}_{Z(j \mathrm{TeV})}$ and $R^{\mathrm{tot}}_{\tt_i/\tt_j}=\sigma^{\mathrm{tot}}_{\tt(i \mathrm{TeV})}/\sigma^{\mathrm{tot}}_{\tt(j \mathrm{TeV})}$}
\item{$R^{\mathrm{tot/fid}}_{\tt/Z}(i\mathrm{TeV})=\sigma^{\mathrm{tot}}_{\tt(i \mathrm{TeV})}/\sigma^{\mathrm{fid}}_{Z(i \mathrm{TeV})}$}
\item{$R^{\mathrm{tot/fid}}_{\tt/Z}(i/j)=[\sigma^{\mathrm{tot}}_{\tt(i \mathrm{TeV})}/\sigma^{\mathrm{fid}}_{Z(i \mathrm{TeV})}]/[\sigma^{\mathrm{tot}}_{\tt(j \mathrm{TeV})}/\sigma^{\mathrm{fid}}_{Z(j \mathrm{TeV})}]$}
\end{itemize}
This study takes advantage of the precision achieved thanks to the cancellation of several detector-related systematic uncertainties to extract information on the strong coupling constant ($\alpha_S$), the top-quark mass ($m_t$), and the PDFs, with a particular sensitivity to the $q/g$ ratio.
The measurements used to evaluate the ratios were previously published by ATLAS~\cite{Aaboud:2016btc,Aad:2014kva,Aad:2015auj,Aaboud:2016pbd}, and only the fiducial cross section measurement of \Zll\ on 3.2~fb$^{-1}$ data collected at $\sqrts=13$~TeV was performed for the first time for Ref.~\cite{Aaboud:2016zpd}.
The fiducial selection used for this \Z-boson fiducial measurement corresponds to the following requirements: $\pTl>25$~GeV, $|\etalep|<2.5$, and $66<\mll/$GeV$<116$.
The theory predictions compared to the data for the two processes are: DYNLLO 1.5 at NNLO in QCD, and FEWZ 3.1 at NLO in electroweak for the fiducial \Z-boson cross section calculation; and Top++ 2.0~\cite{Czakon:2011xx} at NNLO+NNLL in QCD for the total \tt\ cross-section calculation.
The theoretical uncertainties on the \Z-boson cross-sections range between 2.8\% and 3.7\%, and those on the \tt\ cross-section between 5 and 6\%, with the largest component coming from the PDF uncertainties in both cases.
Those on the predicted ratios, instead, range from 0.2\% up to 2.5\%.

The cross section measurements are dominated by systematic uncertainties, mainly due to the luminosity, the beam energy, and the signal modelling.
In the ratios, the correlations between the different uncertainties can be exploited, to reduce the effects of some of them on the result.
The luminosity uncertainty, for example, is correlated across the same year for different processes, and other experimental uncertainties, like the lepton ones, are correlated depending on the method used to evaluate them.

\begin{figure}[h]
\centering
\includegraphics[width=0.31\textwidth]{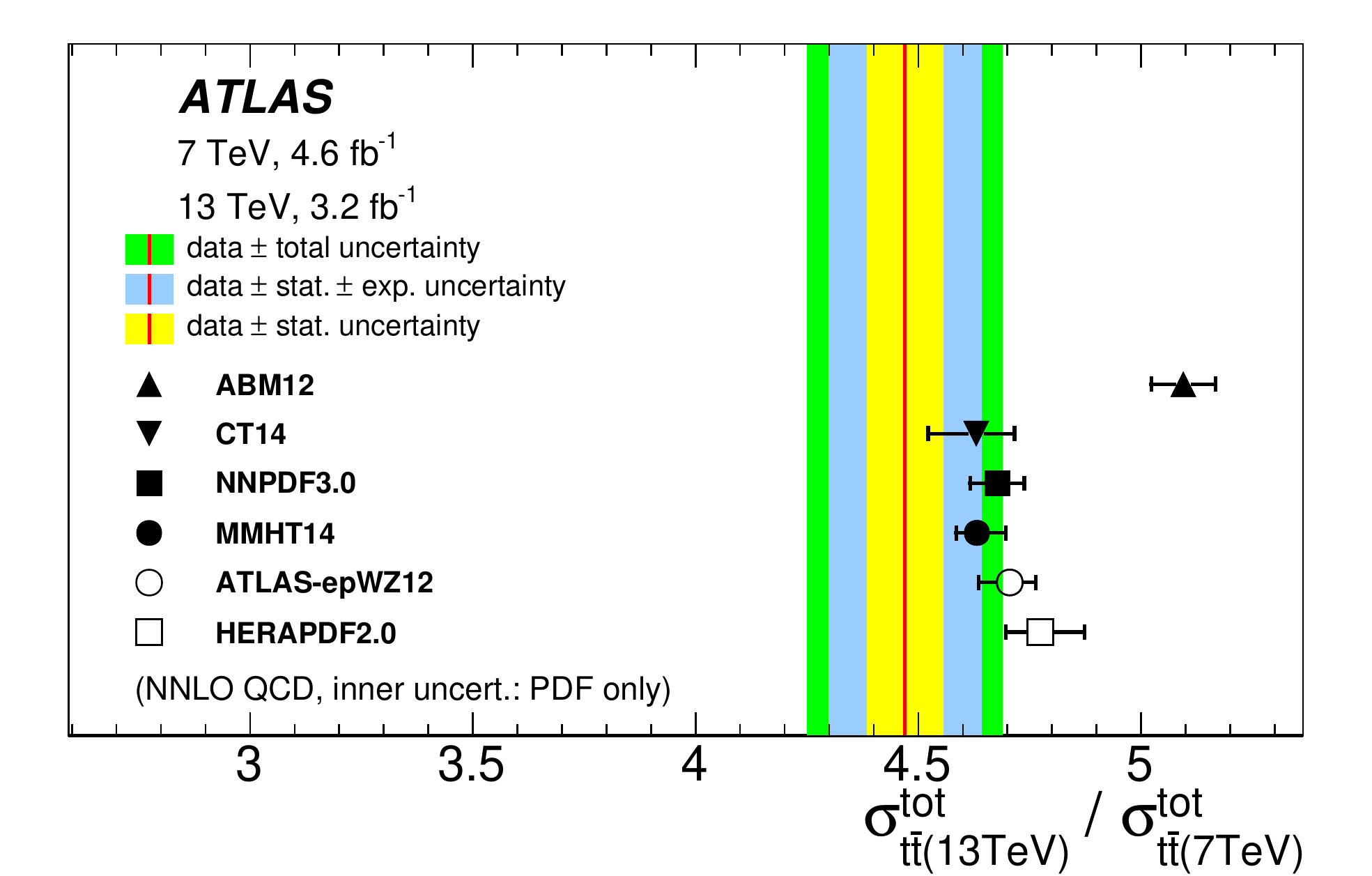}
\includegraphics[width=0.31\textwidth]{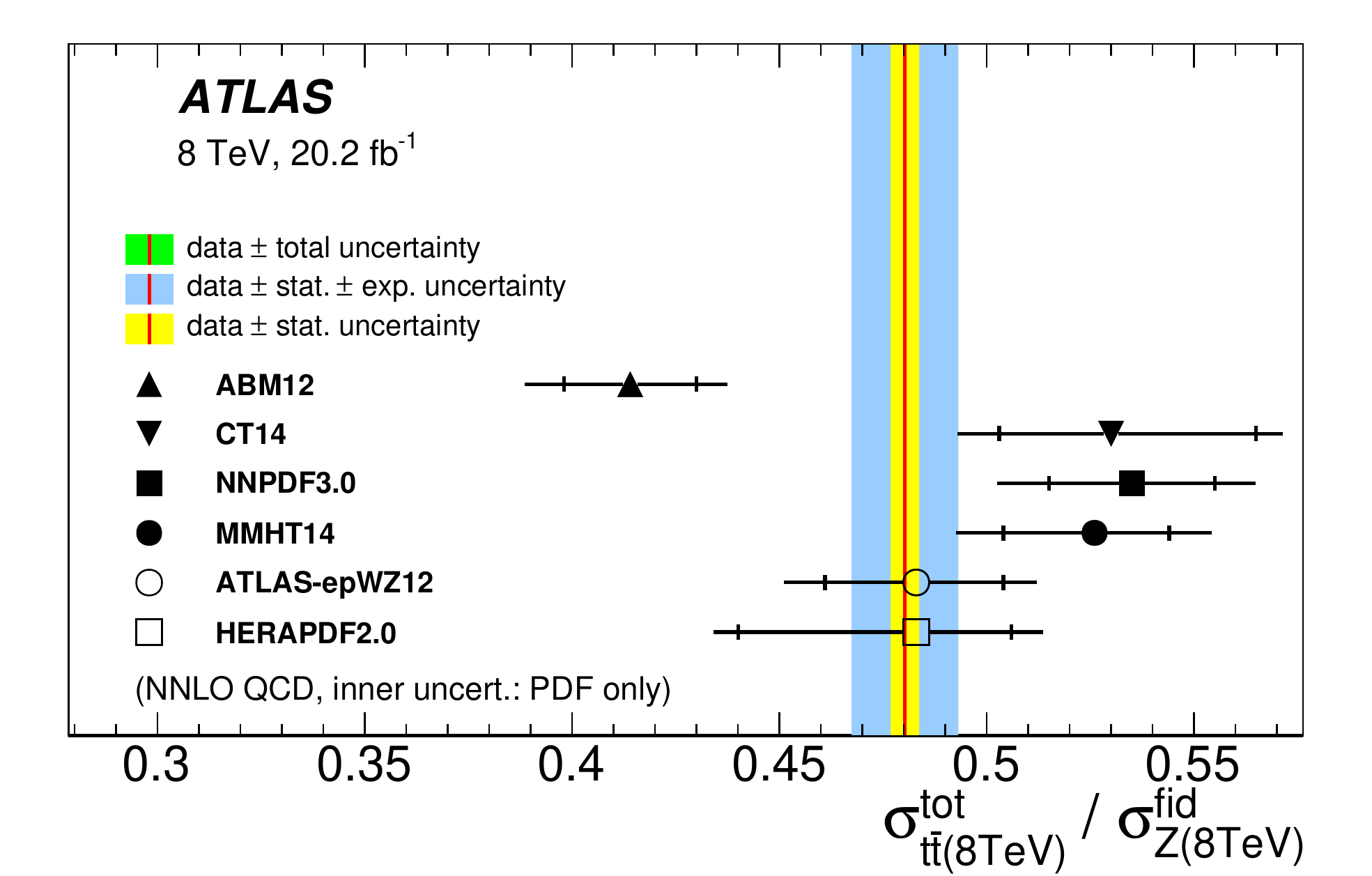}
\includegraphics[width=0.31\textwidth]{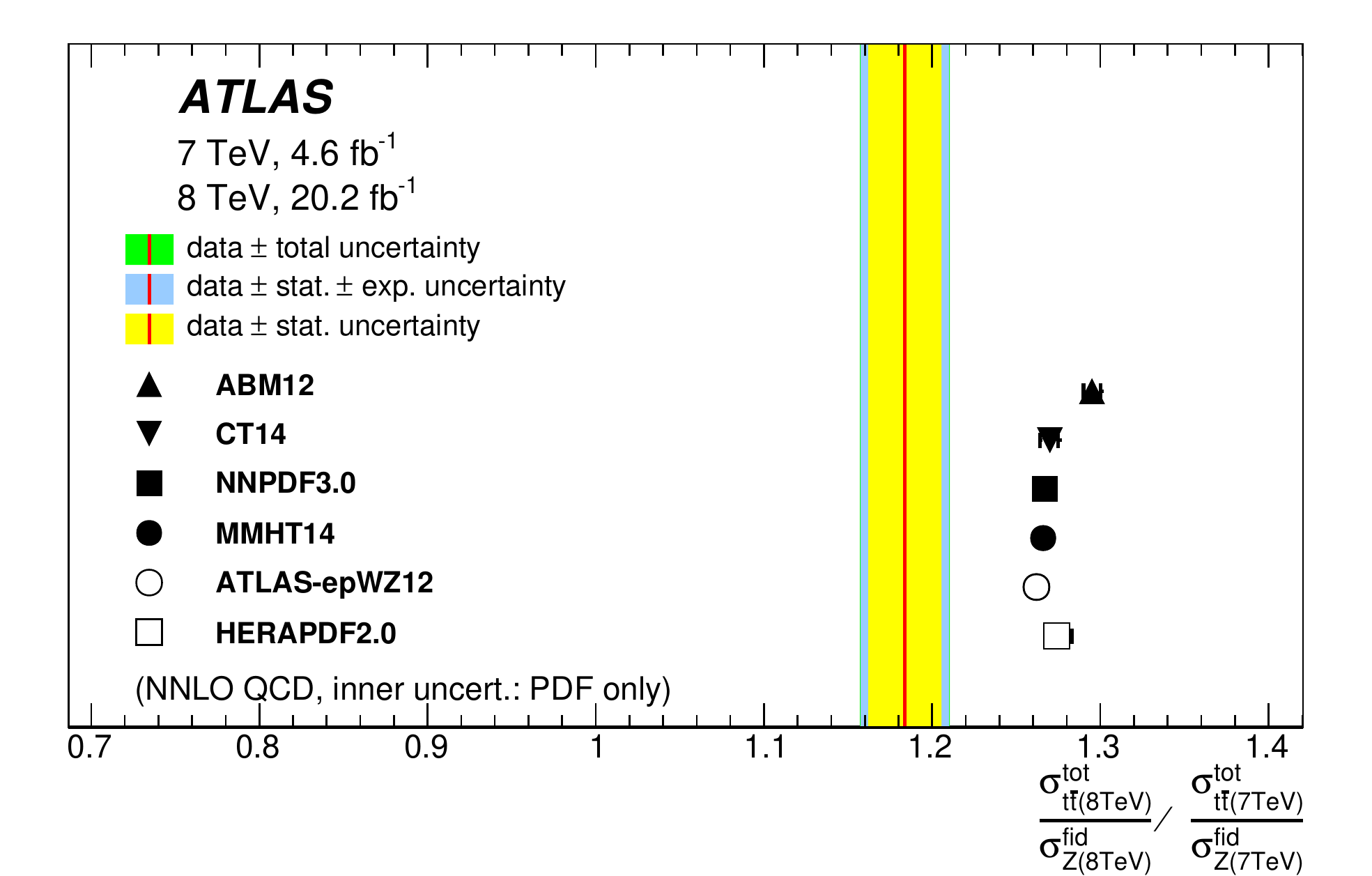}
\caption{The left plot shows the measured ratio between the total \tt\ cross sections at $\sqrts=13,7$~TeV. The central plot shows the ratio of the total \tt\ cross section to the fiducial \Z\ cross-section at $\sqrts=8$~TeV. The right plot shows the measured double ratio of the total \tt\ cross-section to the fiducial one of the \Z-boson at $\sqrts=8,7$~TeV. The theory predictions are given with the corresponding PDF uncertainties shown as inner bars while the outer bars include all other uncertainties added in quadrature. All the plots are taken from Ref.~\cite{Aaboud:2016zpd}.}
\label{fig:xsec_ratios_ttZ}
\end{figure}

The cross section ratios are evaluated in data, and compared to the mentioned predictions using different PDF sets.
The ratios evaluated at different centre-of-mass energies reveal differences mainly due to the different gluon contributions as a function of the Bjorken-$x$ for the \tt\ case, as can be seen in the left plot in Figure~\ref{fig:xsec_ratios_ttZ}, while the agreement is good between data and predictions for the \Z-boson case.
The measurement of the \tt/\Z\ ratios results in a much lower uncertainty on the measurement than on the predictions.
This allows for a discrimination between the latter, which have a rather large spread, due to the different gluon densities and \alfas, as can be observed in the central plot in Figure~\ref{fig:xsec_ratios_ttZ}.
The \tt/\Z\ double ratios at different centre of mass energies exhibit large deviations between the data measurement and the predictions in the case where 8~TeV and 7~TeV are considered, which cannot be fully explained by PDF effects.
The effect can be seen in the rightmost plot in Figure~\ref{fig:xsec_ratios_ttZ}.
The other comparisons between data and predictions show good agreement.

Also for this measurement, studies on the compatibility with different PDF sets, and profiling of these are performed using the measured data. 
The best compatibility is obtained for the ATLAS-epWZ12 PDF set, with $\chisqndf=8.3/6$, 
then profiled using the measured data, obtaining constraints on the light-quark sea distribution functions for $x<0.02$, and on the gluon distribution function at $x\sim0.1$.

\section{Conclusions}

The ATLAS Collaboration, using datasets collected at different centre of mass energies, achieved very high precision in Drell-Yan process measurement, providing tests on the currently available theoretical higher order predictions in QCD and electroweak theories, and valuable information on the proton structure.
The data suggest an enhancement of the strange fractions in the PDFs, and thanks to the measurement of the \tt\ and \Z\ cross section ratios new constraints on the light-quark and gluon fractions were set.
Furthermore, through the high precision measurement of the \W\ and \Z-boson cross-sections, a competitive measurement of the CKM $|V_{cs}|$ matrix element was obtained, and differential studies as a function of the boson rapidity were compared to the available predictions, exploiting the sensitivity to different PDF predictions.


\end{document}